\documentclass[conference]{IEEEtran}

\def \shownames {Show names}

\ifdefined\showchanges
\newcommand{\changed}[1]{\textcolor{blue}{#1}}
\else
\newcommand{\changed}[1]{#1}
\fi

\usepackage{balance}
\usepackage{enumitem}
\usepackage[hidelinks]{hyperref}
\usepackage{letltxmacro}
\usepackage{listings}
\usepackage{longtable}
\usepackage{tabularx}
\usepackage{todonotes}

\LetLtxMacro{\todonote}{\todo}
\renewcommand{\todo}[2][]
{\todonote[inline, caption={#2}, size=\footnotesize, #1]
{\renewcommand{\baselinestretch}{0.5}\selectfont#2\par}}

\newcommand{\materials}{\url{https://github.com/SGX-Cloud}}

\begin{document}

\title{Migrating SGX Enclaves with Persistent State}

\ifdefined\shownames
\author{
\IEEEauthorblockN{
Fritz Alder\IEEEauthorrefmark{1},
Arseny Kurnikov\IEEEauthorrefmark{1}, 
Andrew Paverd\IEEEauthorrefmark{1}, 
N. Asokan\IEEEauthorrefmark{1}}
\IEEEauthorblockA{
\IEEEauthorrefmark{1}
Aalto University, Finland\\
fritz.alder@acm.org
arseny.kurnikov@aalto.fi,
andrew.paverd@ieee.org,
asokan@acm.org}
}
\fi

\maketitle

\ifdefined\showchanges
\raggedbottom
\input{sections/changes}

\IEEEpeerreviewmaketitle

\fi

\begin{abstract}

Hardware-supported security mechanisms like Intel Software Guard Extensions (SGX) provide strong security guarantees, which are particularly relevant in cloud settings.
However, their reliance on physical hardware conflicts with cloud practices, like migration of VMs between physical platforms.
For instance, the SGX trusted execution environment (\emph{enclave}) is bound to a single physical CPU.

Although prior work has proposed an effective mechanism to migrate an enclave's data memory, it overlooks the migration of persistent state, including sealed data and monotonic counters; the former risks data loss whilst the latter undermines the SGX security guarantees.
We show how this can be exploited to mount attacks, and then propose an improved enclave migration approach guaranteeing the consistency of persistent state.
Our software-only approach enables \emph{migratable} sealed data and monotonic counters, maintains all SGX security guarantees, minimizes developer effort, and incurs negligible performance overhead.

\end{abstract}

\section{Introduction}
\label{sec:introduction}

Cloud computing has brought numerous benefits in terms of scalability, elasticity, and efficiency.
However, this outsourcing of computation to potentially untrusted machines has also given rise to various security concerns.
It is widely acknowledged that hardware-supported security mechanisms can be used to address these concerns.
For example, Intel's Software Guard Extensions (SGX) technology enables an application to create trusted execution environments, called \emph{enclaves}, in which security-sensitive data can be stored and processed.
The code and data inside an enclave is protected from all other software on the platform, including the OS and hypervisor.
Using \emph{remote attestation}, an enclave can provide strong assurance to a remote party about the precise software being run inside the enclave.
However, the strong hardware-enforced security guarantees provided by SGX inherently conflict with cloud practices.
For instance, how can a virtual machine (VM) containing one or more SGX enclaves be securely migrated between physical machines?

Previous work by Park et al.~\cite{live-sgx-migration-services} and Gu et al.~\cite{live-sgx-migration-dsn} has considered the challenge of migrating SGX enclaves between physical machines.
Park et al.~\cite{live-sgx-migration-services} first identified the central challenge of migrating the data held securely within an enclave, and suggested that this could be solved with a new hardware instruction.
The state-of-the-art solution by Gu et al.~\cite{live-sgx-migration-dsn} solves this challenge using a software-only approach.
By adding a library inside the enclave, they are able to pause the operation of an enclave and write out the enclave's memory pages, encrypted for the same enclave on the destination machine.
In the destination enclave, the equivalent library reads in and maps these encrypted memory pages, and allows the enclave to resume execution.

However, neither of these previous approaches have considered enclaves that require persistent state, which is stored outside the enclave itself.
Specifically, they do not support migration of enclaves that use \emph{sealed data} and/or \emph{monotonic counters}.
\emph{Sealing} is a feature that allows an enclave to encrypt data with a key known only to the enclave (the \emph{sealing key}), so that the encrypted data can be securely stored in persistent storage outside the enclave.
The sealed data can later be returned to the enclave and decrypted. 
By default, SGX sealing uses authenticated encryption (i.e., AES-GCM), which allows the enclave to detect if the sealed data has been modified externally.
However, in the SGX threat model, the OS and hypervisor are not trusted, and so the enclave has no guarantee that it has received the latest version of the sealed data.
To overcome this challenge, SGX provides each enclave with up to 256 hardware-backed monotonic counters.
An enclave can read and increment these counters, and SGX guarantees that the counter cannot be decremented.
When the enclave seals data, it increments a counter and includes the new counter value in the sealed data.
Thus when the data is unsealed, the enclave can detect if it is not the most recent version by comparing the included value against the hardware counter's current value.
Additionally, monotonic counters can be used for various other application-specific purposes within the enclave (e.g., keeping track of transactions performed by the enclave).

We refer to sealed data and monotonic counters collectively as the \emph{persistent state} of an enclave. 
As we show in Section~\ref{sec:threats}, failure to migrate this persistent state could lead to data loss or attacks against \emph{otherwise secure} systems, such as the recent Teechan payment system~\cite{teechan-bitcoin} or the Hybster state-machine replication protocol~\cite{hybster-eurosys}.

According the SGX Developer Guide~\cite{SGX-dev-guide}, an enclave is destroyed, and its data memory irrecoverably lost whenever:
\begin{itemize}
\item The application closes the enclave; or
\item The application itself is closed or crashes; or
\item The machine is hibernated or shutdown.
\end{itemize}
\noindent
Enclaves should therefore always have the ability to store their important data as persistent state at short notice, whenever any of the above events take place.  
It is reasonable to assume that most real-world enclaves have data that must be persisted (e.g., cryptographic keys or other secrets provisioned to the enclave through remote attestation).
Even if an enclave stores some of its data on a shared storage, or other remote device, so that it can be accessed by the enclave from different physical machines, it still requires local persistent storage for key and counters needed for using that shared storage.
Thus when considering enclave migration, it is critical to consider migration of persistent state.

In this paper, we define a set of requirements to address this challenge (Section~\ref{sec:requirements}) and propose an improved mechanism for migrating enclaves with persistent state (Section~\ref{sec:design}).
We achieve this by introducing a separate \emph{Migration Enclave} on both the source and destination machines.
Our approach focuses only on the migration of persistent state because, as explained above, a well-designed enclave should always have the ability to store its important data as persistent state.
Enclave can thus use this same mechanism to persist any important information before they are migrated.
If there are cases where this is not feasible, we assume the enclave developers will use an additional mechanism, such as that proposed by Gu et al.~\cite{live-sgx-migration-dsn}.
\changed{Given the design of Intel SGX, it is not possible to achieve \emph{transparent} migration of enclaves without requiring hardware modifications.
However, as a first step, we present a software-only approach can be used on existing SGX hardware, and then discuss what would be needed to enable transparent migration.} 
We provide a proof-of-concept implementation, available as open source software (Section~\ref{sec:implementation}), and use this to evaluate the security, usability, and performance of our solution (Section~\ref{sec:evaluation}).

In summary, we claim the following contributions:

\begin{itemize}

\item We motivate the need to migrate persistent state by showing how the omission of state migration could lead to data loss and/or undermine the security guarantees of SGX (Section~\ref{sec:threats}).

\item We define an improved set of security requirements (Section~\ref{sec:requirements}) and propose a new software-only architecture and protocol for securely migrating SGX enclaves with persistent state (Section~\ref{sec:design}).

\item We demonstrate the feasibility of our approach by providing an open-source proof-of-concept implementation\footnote{All software is available at: \materials} (Section~\ref{sec:implementation}), and we use this to evaluate the security, performance, and usability of our approach (Section~\ref{sec:evaluation}). 

\end{itemize}

\section{Preliminaries}
\label{sec:preliminaries}

\subsection{Intel SGX}

\subsubsection{Isolated execution} 
Intel's Software Guard Extensions (SGX) technology is a set of CPU instructions that applications can use to create \emph{enclaves} -- isolated execution environments containing security-sensitive data and the software that operates on this data. 
Once an enclave has been initialized, SGX ensures that the software running within the enclave cannot be modified from outside the enclave.
It also ensures that execution of the enclave's software can only begin from well-defined entry points, known as ECALLs, to avoid software attacks.

\subsubsection{Memory protection} 
The SGX hardware ensures that only code within the enclave can access the enclave's memory, thus protecting it against all other software on the platform, including the OS and/or hypervisor.
Specifically, the enclave's memory is mapped to a special area of physical memory called the Enclave Page Cache (EPC).
When enclave data leaves the physical CPU boundary (e.g., is written to DRAM), it is automatically encrypted to protect against attacks like memory bus snooping.
SGX provides integrity protection and anti-replay protection for this memory, to prevent encrypted memory pages being reverted to earlier versions.

\subsubsection{Enclave and signing identities}
When an enclave is loaded, its software is \emph{measured} to produce the \emph{enclave identity}, also referred to as the MRENCLAVE value.
Specifically, for each memory page within the enclave, the contents and the properties of the page are hashed to create a unique yet deterministic representation of the enclave.
This process will result in the same value on any physical machine.
Each enclave can also be signed by the enclave developer, and the hash of the developer's public key constitutes the enclave's \emph{signing identity}, also referred to as the MRSIGNER value.

\subsubsection{Sealing}
Sealing refers to the process of encrypting data with a key known only to the enclave, so that the sealed data can be stored outside the enclave.
The SGX SDK provides a default sealing function \texttt{sgx\_seal\_data()}, which obtains an enclave- and machine-specific sealing key from the CPU and encrypts data using an authenticated encryption algorithm (i.e., AES-GCM).
Data can be sealed against either the enclave's identity (MRENCLAVE value) or the signing identity (MRSIGNER value)~\cite{Anati2013}. 
Data sealed against MRENCLAVE can only be decrypted by the same enclave, whereas data sealed against MRSIGNER can be unsealed by any enclave signed by the same developer (e.g., to allow upgrades of enclave code).
In both cases, the encryption key (sealing key) is derived from a CPU-specific secret, so data can only be unsealed on the same physical machine on which it was sealed.
SGX sealing thus guarantees the confidentiality and integrity of sealed data, but does not automatically provide roll-back protection.
If required, the enclave developer must check that the correct version of the sealed data has been provided.
Usually this can be achieved using secure monotonic counters.

\subsubsection{Monotonic counters}
With support from the Intel Platform Software~\cite{intel-sgx-sdk-reference}, each SGX enclave has access to up to 256 enclave-specific monotonic counters. 
These counters are maintained by the platform's hardware and firmware (e.g., the Intel Management Engine).
Counters are thus specific to a physical machine, in the same way as sealing keys.
When a monotonic counter is created, Intel Platform Software assigns it a counter \textit{UUID} which consists of a \textit{counter ID} and a \textit{nonce}.
The counter ID uniquely identifies the counter while the nonce ensures that it can only be accessed by the enclave that created it.
Because of this mechanism, it is not possible to destroy a counter and create a new one with the same identifier but lower value on the same physical machine.

\subsubsection{Attestation}

An SGX enclave can use \emph{attestation} to provide strong assurance of its identity to a relying party.
We refer to the attested enclave as the \emph{prover} and the relying party as the \emph{verifier} in the attestation protocol.
The verifier can use the information included in the attestation to authenticate the prover and establish a secure communication channel with the prover.
There are two types of attestation: local and remote.

\emph{Local attestation} allows an SGX enclave to prove its identity and authenticity to another enclave on the same physical machine.
Specifically, the prover enclave uses an SGX hardware instruction to generate a \emph{report} for the verifier enclave.
The report includes the identity of the prover enclave, and may also contain application-specific data.
The CPU generates a message authentication code (MAC) for the report, using a symmetric key available only to the verifier enclave. 
Thus local attestation inherently guarantees that the prover is a genuine SGX enclave running on the same machine as the verifier.
Two local enclaves can establish a secure communication channel by performing mutual local attestations and including key agreement messages in the reports.
Local attestation is also used to enable communication between application enclaves (i.e., enclaves from third-party developers) and \emph{architectural enclaves} provided by Intel.
For example, the Platform Services Enclave is an architectural enclave that allocates and manages the monotonic counters for application enclaves.

\emph{Remote attestation} allows an SGX enclave to prove its identity to a verifier (not necessarily an enclave) on another physical machine.
It also assures the verifier that the prover is a genuine SGX enclave.
In order to achieve this, the prover enclave first performs local attestation with its local Quoting Enclave (QE), another architectural enclave provided by Intel.
The QE creates an SGX \emph{quote} containing the identity of the prover enclave (MRENCLAVE or MRSIGNER) and any application-specific data provided by the prover enclave.
The QE signs the quote using the Enhanced Privacy ID (EPID) scheme~\cite{EPID}, a group signature scheme that allows revocation of compromised components.
This signature on the quote can be verified using the Intel Attestation Service (IAS).\footnote{\url{https://software.intel.com/en-us/blogs/2016/03/09/intel-sgx-epid-provisioning-and-attestation-services}}

\subsection{SGX Virtualization and Migration}

Virtualization is a key enabler of cloud computing, allowing multiple virtual machines (VMs) to be run on a single physical machine.
Typically, a \emph{hypervisor} is used to manage and schedule the different VMs on a machine.
Virtualization aims to provide better resource utilization, improved scalability, and ease of maintenance.
VMs are supposed to be self-contained, disposable and easily migratable from one physical machine to another.
Currently, there is relatively little support for SGX in virtualized environments.
Experimental patches to Xen and KVM managers are available~\cite{sgx-virtualization}, but their functionality is limited to allocating EPC pages to a virtual machine.

Migration is a process of ``moving'' a virtual machine (VM) from one physical machine (source) to another (destination).
In \emph{live migration}, the migration is transparent to the VM~\cite{fast-migration-of-vms}.
The memory pages of the VM are copied from the source machine to the destination machine, and then execution continues from the same place on the destination machine.
If a VM containing an SGX enclave were migrated using existing VM migration techniques, the enclave would not be migrated because the migration process would not be able to access the EPC.
In order to migrate the enclave, an SGX-aware migration mechanism, such as that proposed by Park et al.~\cite{live-sgx-migration-services} or Gu et al.~\cite{live-sgx-migration-dsn} must be used.

However, neither of these SGX-aware migration mechanisms are able to securely migrate enclaves that include persistent state.
If an enclave sealed any data on one physical machine it will not be able to access it after migration, because the sealing key on the destination machine will differ. 
Even though the enclave's identity remains unchanged, the sealing key is derived from the CPU secret, which is unique to each physical machine.
Similarly, any monotonic counters created by the enclave on the source machine will be lost when the enclave migrates, because these are machine-specific.

\section{Threat Model and Attacks}
\label{sec:threats}

In this section we define the threat model for an enclave migration mechanism, and describe potential attacks that could arise if an enclave's persistent state is not migrated.

\subsection{Threat model}
\label{sec:threat_model}

We assume the same threat model as SGX~\cite{SGX-innovative-instructions}, in which the trusted computing base (TCB) for a specific enclave consists only of the SGX hardware and the code within that enclave. 
From the enclave's perspective, all other software on the machine, including the OS and hypervisor, is untrusted.

We thus assume that the adversary has physical access to the machine, privileged access to all software (including OS, and hypervisor), and the ability to monitor and manipulate all network traffic.
As usual, we assume that the adversary is unable to subvert correctly implemented cryptographic primitives and, in general, is unable to subvert the security guarantees of SGX on any single machine.
However, recent research on side-channel attacks against SGX~\cite{controlled-channel-oakland, high-resolution-side-channels-usenix, branch-shadowing-usenix} has shown that in some cases, the latter assumption may not hold.
We evaluate the security of our scheme with respect to side-channel attacks in Section~\ref{sec:evaluation}.

The adversary's goal is to use the migration mechanism to subvert the SGX security guarantees, which would otherwise not be possible.
In particular, the adversary aims to mount either a \emph{fork attack} or a \emph{roll-back attack}, as described below.
\changed{We assume that the adversary's goals do not include denial-of-service attacks, since our strong adversary already has full control over the availability of the physical machine and all enclaves.}
Thus, his motives are not to deny the service but to undermine its security.

\subsection{Fork Attack}
\label{sec:threats-fork}

The objective of a \emph{fork attack} is to create two or more copies of the same enclave with inconsistent state, potentially running on different machines, in order to undermine some application-specific security guarantee.
In this section we consider two recent SGX-based systems that are currently secure but would become insecure if they were made migratable using a mechanism that did not migrate persistent state.

\emph{Teechan}~\cite{teechan-bitcoin} is a framework for establishing full-duplex payment channels between SGX enclaves, in order to support frequently-repeated (micro) payments using blockchain-based cryptocurrencies.
Once two enclaves have established a channel, they can exchange funds in either direction with a single message.
The authors explain that the enclaves can \emph{``persist their state to secondary storage, encrypted under a key and stored with a non-replayable version number from the hardware monotonic counter''}~\cite{teechan-bitcoin}.

The Hybster state-machine replication protocol~\cite{hybster-eurosys} introduces a trusted subsystem, called \emph{TrInX}, that provides a trusted counter service using SGX.
TrInX counters are distinct from SGX hardware monotonic counters. 
The authors correctly assume that \emph{``the execution platform provides a means to prevent undetected replay attacks where an adversary saves the (encrypted) state of a trusted subsystem and starts a new instance using the exact same state to reset the subsystem''}~\cite{hybster-eurosys}.
Although not stated, it can be assumed that this guarantee would be provided using sealing and hardware monotonic counters, as in Teechan.

We assume that in both Teechan and TrInX, any important state information would be persisted before the enclave is terminated.\footnote{In practice, these enclaves might persist their state more frequently (e.g., to mitigate against unexpected crashes), but at a minimum they must do this at least once before enclave termination.}
This would be done by incrementing a hardware counter and sealing the new counter value along with the enclave's state as a version number.
When the enclave is restarted, it would only accept sealed data for which the version number matches the current hardware counter value.

Now suppose either of these enclaves were made migratable using a mechanism that \emph{does not migrate hardware counters}.
A fork attack could proceed as follows:

\begin{enumerate}

\item \label{fork1} \changed{\textbf{Start-stop-restart:} Start the enclave on the source machine and then signal that the application process will be terminated. 
This causes the enclave to request a monotonic counter $c$ (since this is the first use), increment the counter value ($c=1$), and store its keys and important information as persistent state  with the current counter value as a version number ($v=1$).
Restart the application on the source machine using the persistent state.}

\item \label{fork2} \changed{\textbf{Migrate:} Migrate the VM containing this application to the destination machine and continue operation (e.g., make transactions with Teechan and update TrInX counters).
The application may persist its state arbitrarily many times ($v=2,3,4,...$) using the monotonic counter on the destination machine ($c'$).}

\item \label{fork3} \changed{\textbf{Terminate-restart:} Terminate the application process on the source machine, then restart this process (still on the source machine) using the persistent state created in step~1 ($c=v=1$).}

\end{enumerate}

\noindent
This would allow two copies of the enclave to execute concurrently on different machines with inconsistent state, thus undermining the guarantees required by Teechan and Hybster.
The mechanism by Gu et al.~\cite{live-sgx-migration-dsn} only partially mitigates such fork attacks.
In their protocol, once an enclave has been migrated, the enclave on the source machine is prevented from resuming operation by setting a flag that holds all the enclave's worker threads in a perpetual spin lock.
However, the authors do not state whether this flag is stored in persistent storage.
If the flag is not persisted, the above attack will succeed because the flag will be cleared when the source enclave is terminated and resumed (step~3).
Alternatively, if the flag is persisted, the fork attack will be prevented, but this would prevent the same enclave from ever being migrated back to the source machine, since this is indistinguishable from a fork attack.
The latter would place significant constraints on how the cloud operator can manage the migration of VMs, thus reducing the benefits of cloud computing.
Therefore, migration of hardware counters calls for careful design.

\subsection{Roll-back Attack}
\label{sec:threats-rollback}

In addition to fork attacks, the adversary may also be able to mount roll-back attacks under certain circumstances.
As explained above, SGX-based systems like Teechan and TrInX can protect the confidentiality and integrity of their persistent state by encrypting it using a key available only to the enclave.  
On a single machine, this could be achieved using the SGX sealing functionality, but since the sealing key is machine-specific (as explained in Section~\ref{sec:preliminaries}), sealed data cannot be directly migrated with a VM.
The mechanism by Gu et al.~\cite{live-sgx-migration-dsn} does not support migration of sealed data.
This could result in data loss for enclaves that use sealed data.
In order to overcome this, an improved migration mechanism could allow enclaves to seal data under a migratable key, such that it could still be unsealed after migration. 

As an alternative to SGX sealing, an enclave in a cloud environment could request an encryption key from a Key Distribution Center (KDC), such as the AWS Key Management Service\footnote{\url{https://aws.amazon.com/kms/}} and use this to encrypt its persistent state.
The encrypted state could be stored in a specialized (high-availability) storage service outside the VM, e.g., Amazon~S3.\footnote{\url{https://aws.amazon.com/s3/}}

In either case, the enclave would be able to access its persistent (sealed) state after migration.
However, if the migration mechanism does not also migrate the enclave's monotonic counters, this could lead to a roll-back attack as follows:

\begin{enumerate}

\item \changed{\textbf{Start-stop-restart:} \emph{[As in step~\ref{fork1} of the fork attack]}}

\item \changed{\textbf{Continue:} Diverging from the fork attack, continue operation on the \emph{source} machine (e.g., make transactions with Teechan and update TrInX counters).
The application may persist its state arbitrarily many times ($v=2,3,4,...$) using the monotonic counter on the source machine ($c$).}

\item \changed{\textbf{Migrate:} Migrate the VM containing the application to the destination machine and continue operation.}

\item \changed{\textbf{Terminate:} Signal that the application process will be terminated, causing the enclave to persist its state.
Since no counters have yet been created on the destination machine, the enclave requests and increments a new counter ($c'=1$).}

\item \changed{\textbf{Restart:} Restart the application on the destination machine, but provide the original data package from the source machine created in step~1 ($v=1$).
This is accepted by the enclave because the counter value on the destination machine matches the version number in the sealed data ($c'=v=1$).}

\end{enumerate}

\noindent
Thus by abusing the migration process, the adversary can roll-back the state of the enclave's monotonic counters, which would otherwise not be possible.
If it can be performed repeatedly, this roll-back attack could achieve the same result as the fork attack described above.
Both of these attacks undermine the current SGX security guarantees, and could thus have serious consequences for enclaves that use persistent state.

\section{Requirements and Goals}
\label{sec:requirements}

\subsection{Security Requirements}
In order to prevent the attacks described above, the migration mechanism must meet the following requirements:

\begin{enumerate}[label={R\theenumi},leftmargin=*,labelindent=0mm,labelsep=2mm]
  \item \label{R1} \textbf{SGX guarantees:} The migration mechanism must maintain all SGX security guarantees that would be available to an equivalent non-migratable enclave.
  \item \label{R2} \textbf{Controlled migration:} Migration must only be possible if authorized by the machine owner, and the enclave must migrate to the correct destination machine.
  \item \label{R3} \textbf{Fork prevention:} Any fork of persistent state in a migratable enclave must also be possible against the equivalent non-migratable enclave.
  \item \label{R4} \textbf{Roll-back prevention:} Any roll-back attack against persistent state of a migratable enclave must also be possible against the equivalent non-migratable enclave.
\end{enumerate}

\noindent
Requirement~\ref{R1} ensures that the existing SGX security guarantees still apply to migratable enclaves (e.g., isolated execution etc.).
Requirements~\ref{R2} to \ref{R4} are specific to enclave migration, which is not available by default in SGX.
Requirement~\ref{R2} is important to ensure that the adversary cannot use the migration mechanism to take control of an enclave by migrating it to a machine under his control.
For example, this ensures that an enclave can only migrate to another physical machine within the same cloud data center.
Requirements~\ref{R3} and \ref{R4} directly address the fork and roll-back attacks described in the previous section.
Although it is not possible to categorically prevent these attacks for all enclaves (e.g., poorly-designed enclaves that may already be vulnerable without migration), any migration mechanism must not introduce new possibilities for such attacks. 
As explained in Section~\ref{sec:threats}, denial-of-service attacks are not one of the adversary's objectives, and are thus not in scope.

\subsection{Performance Goals}

\noindent\textbf{Enclave performance:} A migratable enclave should not incur a high performance overhead compared to an equivalent non-migratable enclave.
This means that any changes made to the enclave or other software on the machine in order to enable migration should not noticeably degrade the performance of migratable enclaves.

\noindent\textbf{Migration performance:} Ideally, migrating a VM with enclaves should not take significantly longer than migrating an equivalent VM without enclaves.
Even within data centers, the process of copying the VM's entire memory between two machines can take in the order of seconds~\cite{fast-migration-of-vms}.
Any additional time required to migrate the enclave should ideally be at least an order of magnitude lower.

\subsection{Usability Goals}

\noindent\textbf{Developer effort:} Ideally, the effort required from developers in order to make an enclave migratable should be minimal.
It is reasonable to require some input from the enclave developer because the developer must also ensure that the functionality of the enclave is compatible with migration (i.e., that the enclave should be allowed to migrate).

\section{Design}
\label{sec:design}

Based on the requirements, we design and implement a migration framework for SGX enclaves with persistent data.
It consists of two main components: a \emph{Migration Enclave} running on each physical machine, and a \emph{Migration Library} included in each migratable enclave.
We first present an overview of the whole design and then explain the roles of the Migration Enclave and Migration Library in greater detail.

\subsection{Design Overview}
\label{sec:sub-design}
\begin{figure}[t]
	\begin{center}
    \includegraphics[width=1\columnwidth,trim={10mm 0 4mm 5mm},clip]{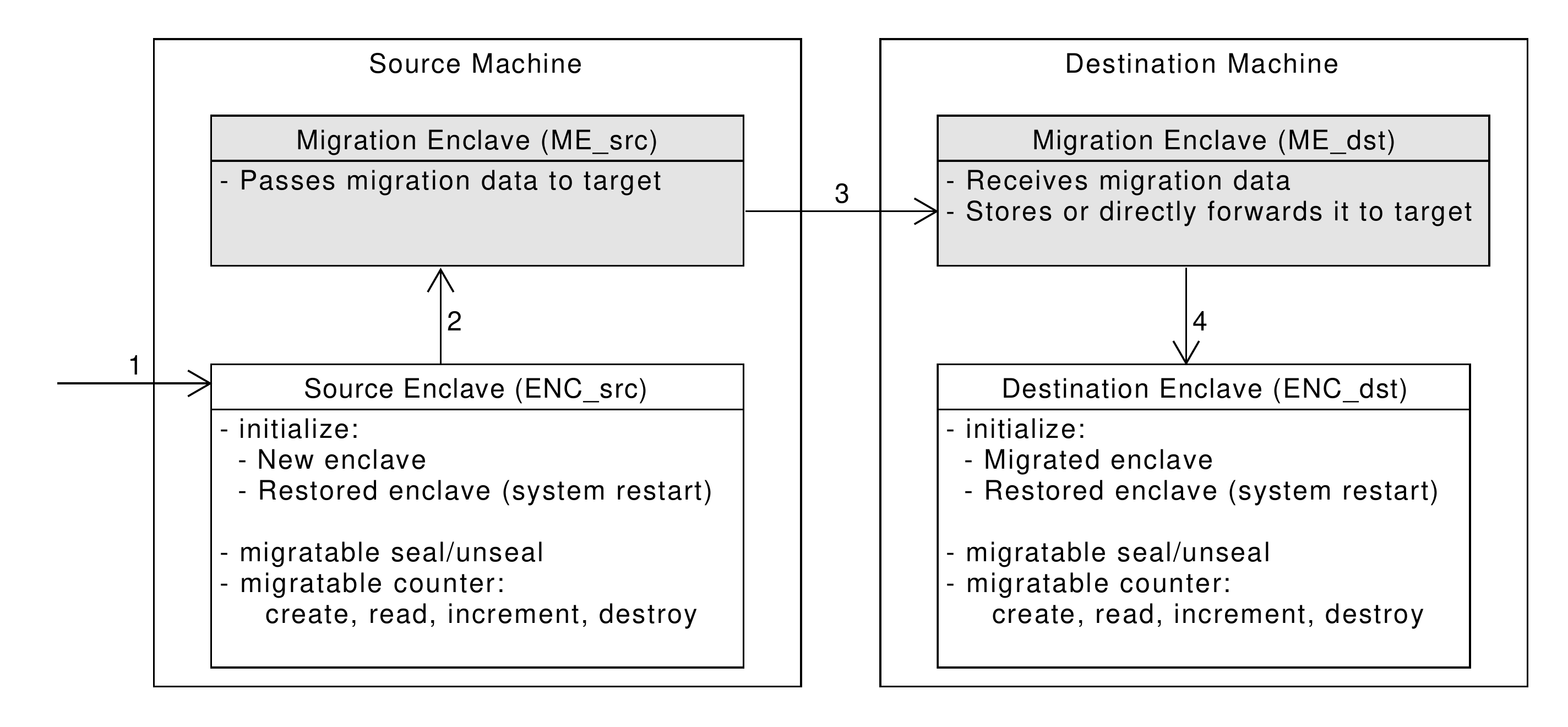}
    \caption{Overview of the migration process.}
    \label{fig:migration_design}
	\end{center}
\end{figure}

The overall design is shown in Figure~\ref{fig:migration_design}.
Each physical machine has a single Migration Enclave, which runs in a separate non-migratable VM.
This is compatible with a typical cloud environment in which each physical machine would have a non-migratable management VM.
The Migration Enclave is responsible for managing the migration process of enclaves.
The enclave developer includes and uses the Migration Library in each migratable enclave.
This library provides migratable versions of the sealing and monotonic counter functions from the SGX API.
It also performs local attestation of the Migration Enclave and communicates with it during the migration process.
The migration process consists of the following steps:

\begin{enumerate}

\item The application notifies the enclave that it will migrate to a new machine. 

\item The Migration Library locally attests the Migration Enclave and sends the data required for the migration. 
At the same time the Migration Library prevents further operations of the enclave. 

\item After the Migration Enclave receives all the data from the library, it performs a mutual remote attestation with the Migration Enclave on the destination machine.
After attesting and establishing a secure channel, the Migration Enclaves authenticate each other to verify that both machines are authorized machines of the same cloud provider.

\item On the destination machine, the Migration Enclave checks that the local destination enclave matches the source enclave and, if so, sends the migration data to the destination enclave's Migration Library to complete the migration.

\end{enumerate}

\noindent
The fact that in our protocol the application has to initiate the migration is neither unrealistic nor a security issue. 
As described in Section~\ref{sec:threats}, denial-of-service attacks are not in scope since SGX does not provide any availability guarantees for an enclave. 
A malicious application, OS, or hypervisor can always prevent enclave migration. 
Our protocol's main goal is to ensure that when a migration does happen, the SGX security guarantees are not violated and the additional requirements defined in Section~\ref{sec:requirements} are met.

\subsection{Migration Enclave}
\label{sec:design-me}
The Migration Enclave is responsible for managing the migration process. 
As described in Section~\ref{sec:sub-design}, its main operations are locally attesting the application enclaves and remotely attesting and authenticating other Migration Enclaves.
To verify the identities of Migration Enclaves running on remote machines, each Migration Enclave is set up by the cloud provider during a secure setup phase.
This setup phase can happen, for example, during the setup of the Management VM that contains the Migration Enclave.
In practice, the setup phase could provide the Migration Enclaves with a key or a certificate from an operator of the data center or by a server owner.
The goal of this setup phase is to ensure that enclaves are only migrated inside the data center or to other trusted servers (Requirement~\ref{R2}). 
The authentication performed by the Migration Enclaves can also be used to limit the migration of enclaves to a certain subset of servers, for example to achieve regulatory compliance.

After the setup phase, the Migration Enclave can be used by any application enclave on the machine by performing a local attestation. 
After the attestation, the Migration Enclave has the guarantee that its communication peer resides on the same physical machine and is safe to migrate.
There are no further checks required from the application enclave as the Migration Enclave guarantees that the application enclave will be migrated to an identical enclave on the destination machine.
\changed{After performing the local attestation and receiving the migration data, the Migration Enclave executes a mutual remote attestation with the corresponding Migration Enclave on the destination machine.
If required, the Migration Enclaves then exchange signatures on the transcript of the attestation protocol, using the keys provisioned by the data center operator, to ensure that they are running in the same data center.
Finally, the migration data is sent via the secure channel.}

For incoming migrations, the Migration Enclave performs a remote attestation, verifies the identity of the remote Migration Enclave, and receives the migration data. 
It then either notifies the local application enclave of incoming data if it is already running, or stores the data temporarily until the local enclave has been started and locally attests to the Migration Enclave to securely transfer the incoming migration data.

\subsection{Migration Library}
\label{sec:design-ml}

As discussed in Section~\ref{sec:preliminaries} and Section~\ref{sec:threats}, it is not sufficient to simply move an SGX enclave from one physical machine to another. 
The purpose of the Migration Library is to provide analogues of specific SGX primitives in order to enable migration whilst providing the same security guarantees.
\changed{Note that the Migration Library and the application enclave that utilizes the library reside in the same protection domain.}
This means that they both trust each other fully as the Migration Library can only protect an enclave that is cooperating and the developer of the application enclave is assumed to have reviewed the Migration Library code before utilizing it in his enclave code.

We identify two main primitives that should be substituted by migratable counterparts: sealing and monotonic counters. 
The sealing key derived from the CPU secret is not suited for encrypting migratable secrets because it is machine-specific.
Thus, our library provides its own implementation of the SGX sealing and unsealing functions by generating a Migration Sealing Key (MSK) and using the MSK for all sealing operations.
\changed{The MSK itself is sealed with the enclave's own sealing key and stored locally.
It is reloaded and unsealed by the Migration Library each time the enclave is restarted.}

Monotonic counters are provided by the Intel Platform Services Software, which includes a system enclave that performs specific tasks such as creating, reading, incrementing, and deleting monotonic counters of an application enclave. 
The Platform Service guarantees that over the lifetime of a counter, it can never be decreased and allows application enclaves to use the monotonic counters for various tasks, such as rollback protection. 
When migrating a monotonic counter, the migration protocol needs to ensure two invariants: 

\begin{itemize}

\item The monotonic counter must be rendered unusable on the source machine in order to prevent fork attacks, as described in Section~\ref{sec:threats-fork}, and thus fulfil Requirement~\ref{R3}.

\item The monotonic counter must be re-created with the same value on the destination machine in order to prevent roll-back attacks, as described in Section~\ref{sec:threats-rollback}, and thus fulfil Requirement~\ref{R4}

\end{itemize}

The Migration Library ensures both of these when migrating a monotonic counter by sending the counter values in the migration data and increasing the counter to its original value before the enclave can continue the execution.
Before the migration data is sent to the Migration Enclave, the Migration Library also commands the Platform Service to delete all counters managed by the library on the source machine.
This prevents fork attacks that aim to restart the Migration Library before the migration has completed but after the counters have been set up on the destination machine.

In addition to providing migratable versions of sealing and monotonic counters, the Migration Library also performs the local attestation of the Migration Enclave and communicates with it to start a migration or to receive a migration on startup via a secured channel.

\subsection{The Migration Process}
\label{sec:design-process}

\begin{figure}[t]
	\begin{center}
    \includegraphics[trim={10mm 25mm 10mm 15mm},clip,width=1\columnwidth]{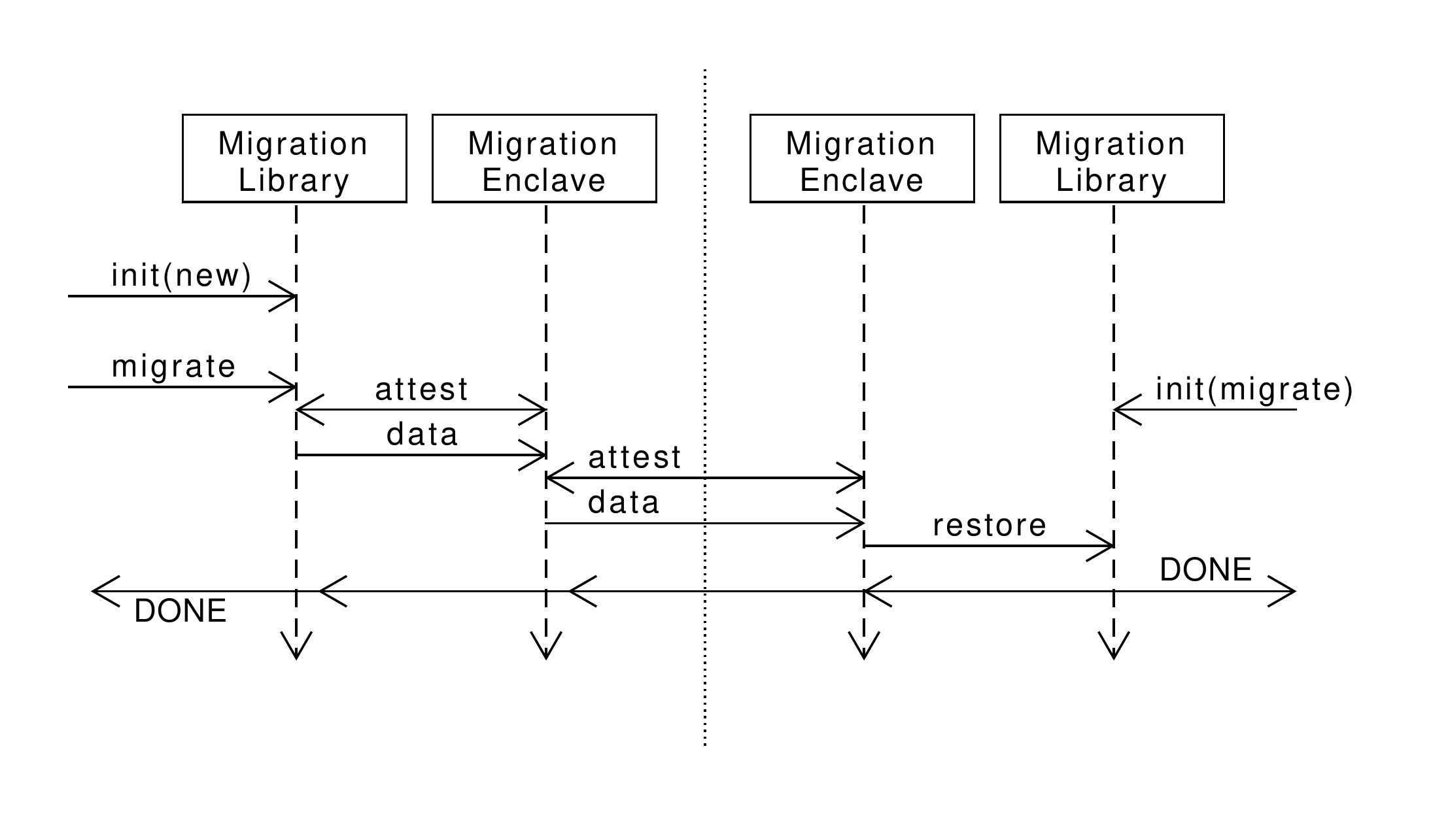}
    \caption{Migration protocol diagram}
    \label{fig:migration_protocol}
	\end{center}
\end{figure}

Figure~\ref{fig:migration_protocol} shows the migration process as an interaction between the Migration Libraries and the Migration Enclaves.
The untrusted parts of the applications are not shown in this figure, but all interaction between the enclaves takes place via untrusted channels.
Therefore, the enclaves establish their own secure channels between one another using attestation.

After initializing the Migration Library, a migration can be requested by the application (e.g., once the application has been notified that it will be migrated).
The Migration Library on the source machine contacts its local Migration Enclave, sets up an encrypted channel based on local attestation, and sends the data to migrate.
This data includes the MSK, the current counter values, and the address of the destination machine. 
After receiving the migration data, the Migration Enclave on the source machine contacts the remote Migration Enclave on the destination machine and sets up an encrypted channel using remote attestation.
This remote attestation includes verifying the integrity of the Migration Enclaves and checking whether they belong to the same cloud provider. 
Next, the data is sent to the Migration Enclave on the destination machine, where it is forwarded to a local Migration Library in the destination enclave.
Either the local Migration Library has already been started and is awaiting an incoming migration, or the Migration Enclave stores the data temporarily until the Migration Library contacts the Migration Enclave.
After sending the migration data to the correct Migration Library, the data is processed and the Migration Library sends back a confirmation of a successful migration.
\changed{This confirmation is sent back to the source machine, which can then safely delete the migration data.
If there is an error during the migration process, the migration data remains in the Migration Enclave on the source machine until the error is resolved or another destination machine is selected to complete the migration process.}

\section{Implementation}
\label{sec:implementation}

\changed{We have implemented the Migration Library and the Migration Enclave and made these available as open source software.\footnote{All software is available at: \materials}
Data center operators can also create and deploy customized versions of the Migration Enclave in order to provision their own policies and certificates.
Similarly, enclave providers can customize the migration library e.g., in order to enforce migration policies.}

\subsection{Migration Enclave}
\label{sec:implementation-me}
As described in Section \ref{sec:design-me}, the Migration Enclave serves as a central entity to handle all migrations of one machine. 
For outgoing migrations, it ensures that the remote Migration Enclave attests with a trusted version of the Migration Enclave code and can authenticate as belonging to the same cloud provider.
This ensures that enclaves can only be migrated to an authorized destination machine, e.g., located within the same data center. 
For incoming migrations, the Migration Enclave ensures that the migration data is only sent to an
enclave that attests with exactly the same version as the source enclave when sending the data. 
We now describe specifics of our local and remote attestation.

The local attestation is kept simple as the Migration Library locally attests to the Migration Enclave and in doing so creates an encrypted channel to the Migration Enclave.
Establishing a secure channel is based on Diffie-Hellman key exchange protocol.
This channel is opened when the Migration Library initializes itself and can be used over the lifetime of the enclave to send and also to receive migrations.
During the local attestation, the Migration Enclave stores the MRENCLAVE value of the calling enclave as it is contained in the REPORT that the local enclave sends to the Migration Enclave during the
attestation. 
This MRENCLAVE value will then be used for incoming and outgoing migrations to match the migration data to the recipient.
On an outgoing migration, the MRENCLAVE value is appended to the migration data of the enclave before sending it to the destination Migration Enclave and on an incoming migration, the migration data is matched to the local enclave that has the same MRENCLAVE value. 
If there is no matching enclave running on the machine for an incoming migration, the migration data will be stored until an enclave with the matching MRENCLAVE value performs a local attestation.  

The remote attestation has more requirements to be checked by the Migration Enclave.
Here, the Migration Enclave is not only interested in checking the actual MRENCLAVE value of its peer but it also has to authenticate its peer.
The first check is trivial as it is contained in the attestation report structure that is used during the attestation.
The Migration Enclave simply checks this MRENCLAVE value and aborts the attestation process if the peer enclave does not have the same MRENCLAVE value as itself.
After establishing the secure channel via remote attestation, the Migration Enclaves can perform the second check and authenticate each other.

\subsection{Migration Library}
\label{sec:implementation-lib}
The migration process has to be started from outside of the enclave, i.e., from the untrusted part of the application.
Listing \ref{lst:enclave} shows the functionality that the Migration Library provides to this untrusted part which is the migration operation and one initialization operation that must be performed every time the enclave is loaded.  
We now describe the migratable sealing and the migratable counter functionality, and then the persistent data that is used in the initialization operation.

\begin{lstlisting}[caption={Migration Library interface for the untrusted part of the application}, captionpos=b, float, frame=tblr, label=lst:enclave]
migration_init(p_data_buffer, init_state, ME_address);
migration_start(destination_address);
\end{lstlisting}

\noindent\textbf{Sealing:}
The migratable version of the sealing functions are straightforward to implement.
Instead of using the built-in sealing functions by Intel SGX, the library generates a Migration Sealing Key (MSK) once for every enclave and uses that for the sealing functions throughout all migrations.
When migrating, the key is transferred to the new Migration Library where it can be used again.
Without re-encryption, the process of migrating the sealed data is constant-time for transferring the key and then linear for transferring the actual sealed data.
This process is secure because the MSK is only ever transferred to trusted Migration Enclaves and from there to the exact same enclave that generated the key.
Thus, the MSK never leaves the trusted environment in an unencrypted form.
From the point of view of the enclave developer, instead of using the standard sealing functions, he now has to use the migratable sealing functions which are identical to the standard functions in terms of parameters required and results returned.
Listing~\ref{lst:enclave-api} shows these new migratable function parameters.
If a developer decides that some data should not be migratable it is still possible to use the native SGX sealing functions to store that data.

\noindent\textbf{Monotonic counters:}
One approach to migrate a counter is for the source enclave to transfer the current counter value to the destination enclave and have the latter create a new counter 
and increment it until the counter value reaches the transferred value. 
However, this will incur significant performance overhead because monotonic counter operations are usually rate-limited. 
Instead, our implementation uses a \emph{counter offset}, which is initialized to zero. 
This offset is added to the \emph{current counter value} to compute the \emph{effective counter value}. 
When an enclave is started for the first time, the counter offset is initialized to zero and the current counter value always equals the effective counter value until the enclave is migrated.
On migration, the effective counter value of the source enclave is sent to the destination enclave where it is set as the new counter offset.
The current counter value on the destination enclave is initialized to zero right after the incoming migration.
\changed{The counter offset remains unchanged between migrations and is sealed and stored locally.}
This design optimizes performance because the processing time of a counter during migration is constant, regardless of the counter value.

If an enclave is malicious it can modify the offset values in the library but we assume that the enclave is trusted, because a malicious enclave could simply lie about counter values or ignore them even when using native SGX counters.
However, the system is potentially open to rollback attacks where an attacker provides old offset values to the Migration Library. 
\changed{To mitigate this rollback attack, when migrating counters to a new destination machine, the used monotonic counters are deleted on the source machine by calling the \texttt{sgx\_destroy\_monotonic\_counter} function.
The process does not proceed until it receives the \texttt{SGX\_SUCCESS} return code, which indicates that the counter has been successfully deleted.}
Only then, the Migration Library sends the effective counter values and the list of active counters in the migration data (see Table \ref{tab:migration-data}) to the  Migration Enclave. 
This ensures that the monotonic counters cannot be used in a rollback attack as Intel SGX ensures that deleted counters cannot be accessed again.
The migratable counter functions in the library ensure that an error is properly thrown if a counter does not exist, no matter what the value of the stored counter offset.

\begin{lstlisting}[caption={Migration Library enclave interface for the application enclave},captionpos=b, float, frame=tblr, label=lst:enclave-api]
sgx_seal_migratable_data(
        additional_MACtext_length, p_additional_MACtext,
        text2encrypt_length, p_text2encrypt,
        sealed_data_size, p_sealed_data);
sgx_unseal_migratable_data(p_sealed_data,
        p_additional_MACtext, p_additional_MACtext_length,
        p_decrypted_text, p_decrypted_text_length);
sgx_create_migratable_counter(p_counter_id, 
        p_counter_value);
sgx_destroy_migratable_counter(counter_id);
sgx_increment_migratable_counter(counter_id, 
        p_counter_value);
sgx_read_migratable_counter(counter_id, p_counter_value);
\end{lstlisting}

As the Migration Library has to perform an extra addition operation on all monotonic counter functions in order to calculate the effective counter value, the Migration Library provides wrapped functions of the standard monotonic counter operations.
In doing this, it assigns the monotonic counters an internal \emph{counter\_id} which can be used to access the counter instead of requiring the Intel SGX UUID of the actual monotonic counter.
Using solely the counter id instead of the Intel SGX UUIDs of the counters is not a security threat as the Migration Library fully trusts the application enclave.
For the enclave developer that integrates the Migration Library this means that instead of storing the UUIDs of the counters himself, he now only has to store the id that the Migration Library assigned to the migratable counter. 
The Migration Library will then handle the internal access to the Intel SGX counter and add the migratable offset before returning the effective counter value to the application enclave.
Listing~\ref{lst:enclave-api} shows the migratable versions of the monotonic counter operations. 
Instead of the standard monotonic counter functions that require a SGX monotonic counter structure to be passed to the function, the migratable version only requires a counter\_id that is assigned and returned on creation of the counter.
This change from the UUID counter structure to the counter\_id number is the only change in the function parameters compared to the standard Intel SGX monotonic counter functions.
\changed{Note that the Migration Library does not require its own monotonic counters and as such the application enclave does not lose any monotonic counter capabilities by utilizing the Migration Library.}
However, because it only wraps the Intel SGX monotonic counters and does not replace them internally, the Migration Library is still limited to the same 256 monotonic counters that are the standard limit for an Intel SGX enclave.

\begin{table}[!t]
\renewcommand{\arraystretch}{1.3}
\caption{Datastructure of the Migrated Data}
\label{tab:migration-data}
\centering
\begin{tabular}{|l|c|c|}
\hline
\bfseries Name & \bfseries Type & \bfseries Description\\
\hline\hline
counters active&bool[256]&Shows used counters\\
counter values&uint32[256]&Used as next offset\\
MSK&128bit SGX key&Used by migratable seal\\
\hline
\end{tabular}
\end{table}

\noindent\textbf{Persistent data:}
One downside of using an MSK and wrapping the monotonic counters for the application enclave is that the Migration Library has a need for persistent data that is stored on the physical machine and then reloaded with every enclave restart.  
This means that the Migration Library needs to be initialized once for the lifetime of a migratable enclave (for key generation) and then every time this enclave is
started (for reloading the key and counter offsets).  
We solve this by handing the data in a sealed data blob over to the untrusted part of the application to store it on the machine. 
Whenever the enclave is started, the Migration Library has to be initialized by the untrusted part of the application with this sealed data buffer.
Table~\ref{tab:library-data} shows the data that is stored in this initialization buffer.
In addition to the MSK and counter arrays, the buffer also stores a flag that identifies whether the enclave has already been migrated.
If this flag is active on initialization, the library will refuse to operate.

\begin{table}[!t]
\renewcommand{\arraystretch}{1.3}
\caption{Datastructure of the Migration Library Internals}
\label{tab:library-data}
\centering
\begin{tabular}{|l|c|c|}
\hline
\bfseries Name & \bfseries Type & \bfseries Description\\
\hline\hline
frozen&uint8&Freeze flag for migration\\
counters active&bool[256]&Shows used counters\\
counter uuids&SGX counter[256]&UUIDs of the SGX counters\\
counter offsets&uint32[256]&Offsets of the counters\\
MSK&128bit SGX key&Used by migratable seal\\
\hline
\end{tabular}
\end{table}

\subsection{Virtualization and Intel Platform Services}
Application enclaves need to contact Intel Platform Services enclaves for various reasons.
In particular, when performing remote attestation the quote is signed by the Quoting Enclave that is part of Platform Services. 
Current SGX patches for KVM modify the system such that enclaves can only run inside VMs.
The Platform Services, on the other hand, relies on a particular piece of hardware (i.e. a PCI bus connected device) to be available.
So this device must be assigned to the specific VM that runs the Platform Services. 
This corresponds well with setting up a management VM on every host. 
It can contain both the Platform Services enclaves and the Migration Enclave.

The SGX SDK uses Unix sockets to communicate between application enclaves and Platform Services enclaves.
Since these are not directly accessible from outside the VM, we introduce two proxies to enable this communication.
One proxy is in the management VM and listens on a TCP socket for incoming connection to pass them on to the Platform Services Unix socket.
The other proxy is running in other VMs. 
It opens the Unix socket for local enclaves to connect to and proxies these connections to the TCP socket inside the management VM.
Original Unix communication is opened to eavesdropping by the untrusted operating system, hence introducing two proxies does not affect the security guarantees.

\section{Evaluation}
\label{sec:evaluation}
In this section, we describe the evaluation of our scheme in terms of security, performance, and usability.

\subsection{Security Evaluation}
\label{sec:evaluation-security}
We evaluate the security of our solution in terms of the four Security Requirements defined in Section~\ref{sec:requirements}. 

\noindent\textbf{SGX guarantees:}
Our library and migration protocols provide the same security guarantees as a stand-alone non-migratable SGX enclave. 
Firstly, the migratable versions of the sealing and monotonic counter operations provide the same security as the standard functions.
The migratable sealing is done with the same encryption method as the standard sealing and the MSK is itself sealed with the standard sealing functions.
The migratable counter operations are also implemented using the standard monotonic counter functions from the Intel SDK, but also add the counter offset to the active counter value before returning.
Secondly, all communication between enclaves is encrypted with symmetric keys established through a Diffie-Hellman key agreement protocol bound to the attestation process.
This ensures that the enclave's migratable sealing key is always protected and is only processed by the Migration Enclaves, which are trusted.
Thirdly, the Migration Enclave on the destination machine uses local attestation to ensure that only an enclave with the same identity as the source enclave can receive the migrated data.
This prevents an attacker from setting up a malicious enclave to receive the migrated enclave data.

The inclusion and use of our Migration Library does not increase the attack surface for side-channel attacks against the application enclave.
The only migratable operation involving secret data is the new migratable sealing function.
However, this function follows the same design as the standard \texttt{sgx\_seal\_data} function (i.e., using the AES-NI hardware instructions to encrypt the secret data), so it is no more vulnerable to side-channel attacks than the standard function.
We therefore conclude that our solution achieves Security Requirement~\ref{R1}.

\noindent\textbf{Controlled migration:}
The identity of the Migration Enclave is verified during the local attestation process before performing a migration.
After that, the identity of the Migration Enclave on the destination machine is verified and authenticated by the Migration Enclave on the source machine. 
This ensures that the destination Migration Enclave has a valid identity and is running on an authorized machine.
Finally, the Migration Enclave on the destination machine only hands over the data to a destination enclave that has the same identity as the source enclave.
We therefore conclude that our solution achieves Security Requirement~\ref{R2}.

\noindent\textbf{Fork prevention:}
Using our scheme, it is not possible for an enclave to be active on two different machines with inconsistent data.
This is ensured by the Migration Enclave which only sends the data to one destination machine and from there to one destination enclave.
Our scheme prevents the type of fork attack described in Section~\ref{sec:threats-fork} because our Migration Library deletes the SGX monotonic counters on the source machine before sending the migration data to the Migration Enclave. 
This is possible because we also store the counters' UUIDs in the persistent data of the Migration Library
If the Migration Enclave is given obsolete persistent data, the monotonic counters are unusable as they would have been deleted by the SDK, even if the Migration Library might not detect that the data is obsolete. 
Hence, trying to access the counter that was initialized from the wrong offset value would result in an error, no matter what the offset value is.
Since we assume that the enclave code is not malicious, it can detect this situation by receiving a monotonic counter error and thus the security is not violated.
Thus our approach fulfils Security Requirement~\ref{R3}.

\noindent\textbf{Roll-back prevention:}
On any given machine, roll-back attacks are prevented using SGX counters (as usual).
Roll-back attacks that could arise from migration (as described in Section~\ref{sec:threats-rollback}) are prevented by migrating the counter values to the destination enclave. 
This means that an enclave's persistent state cannot be rolled back to an earlier version, and thus our scheme fulfils Security Requirement~\ref{R4}.

\noindent\changed{\textbf{Software TCB size:}
As a software-only solution, our Migration Enclave and Library necessarily increase the size of the software Trusted Computing Base (TCB).
However, our Migration Enclave and Library consist of ~217 and ~940 lines of code respectively (excluding the SGX trusted libraries), which is feasible to audit.}

\subsection{Performance Evaluation}
\label{sec:evaluation-performance}

We measured the performance of the migratable alternatives of sealing and monotonic counters, as well as the initialization operations performed when our Migration Library is started for the first time, or subsequently restarted.

All these measurements were obtained by measuring the time of an ECALL for each operation. 
For this we started the enclave, measured the initialization of a new library buffer, restarted the enclave, and measured the other ECALLs. 
After this, we shut down the enclave. 
We repeated this process 1000 times and compared the results with the baseline performance when standard SGX primitives were used for the counter and sealing operations.
Obviously, there is no baseline for the initialization of the library as this step is not required without the Migration Library.
We plot the average of all our results together with error bars that show a 99\% mean confidence interval i.e., the true mean value is within the confidence interval bar with 99\% probability.
We also used a 1-tailed t-test to check if the differences are statistically significant.

\begin{figure}[t]
	\begin{center}
    \includegraphics[width=1\columnwidth,trim={0 0 10mm 12mm},clip]{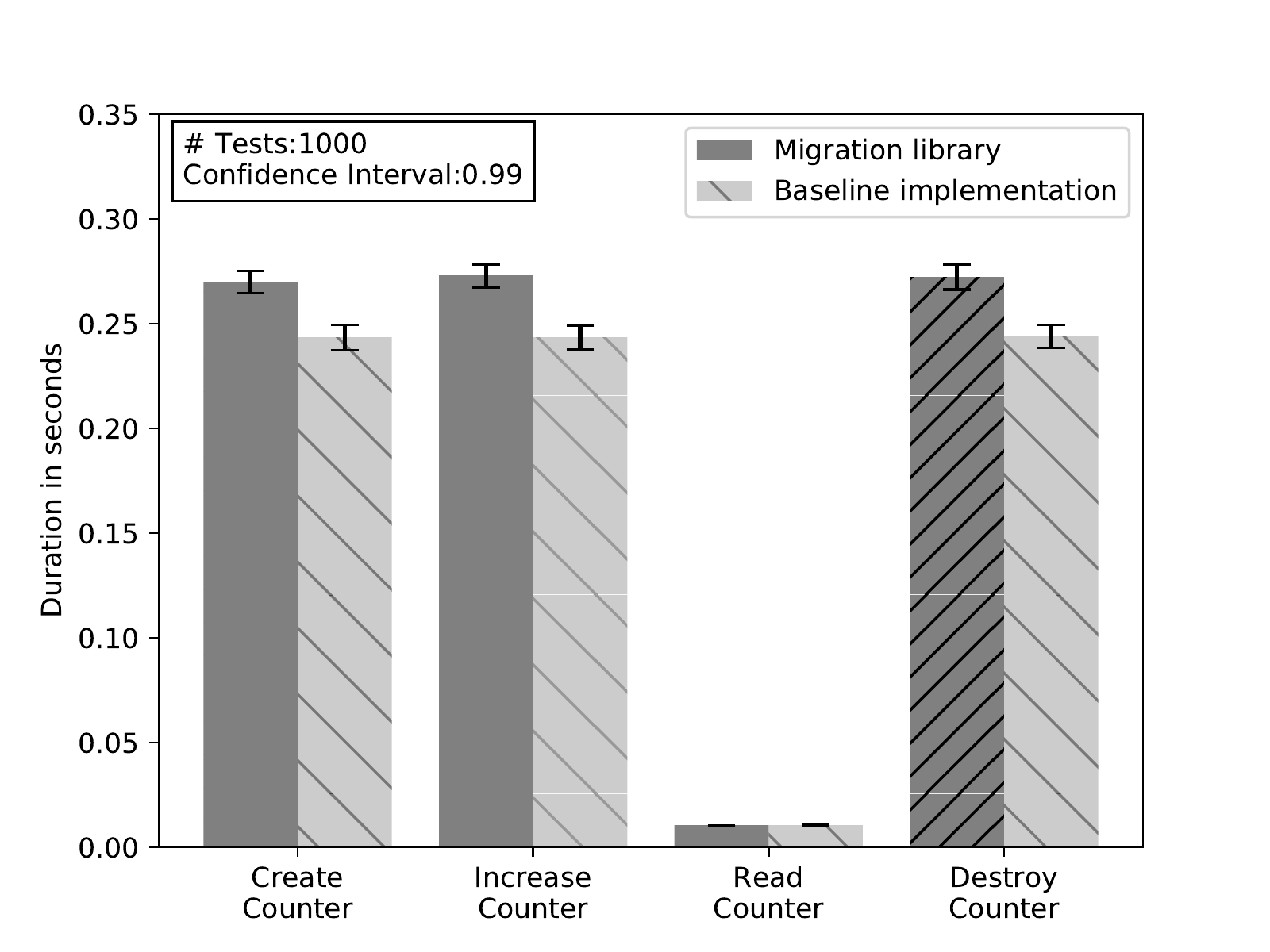}
    \caption{Average duration of counter operations.}
    \label{fig:perf-counters}
	\end{center}
\end{figure}

Figure~\ref{fig:perf-counters} shows the results of the monotonic counter operations.
The migratable versions of the counters introduce a small overhead of at most 12.3\%. 
This is a consequence of the additional array operations performed by the library in order to properly wrap the monotonic counter for the enclave.
For creating and destroying a counter, this means an additional sealing of the internal data buffer that stores the counter data, and for incrementing a counter this means performing additional
checks to prevent an integer overflow due to the offset. 
The increment operation incurs an average overhead of 12.3\% (statistically significant, $p\simeq0$) for migratable counters, whereas the read operation has no statistically significant overhead ($p\simeq0.12$).
Since reading and incrementing counters are the most frequent operations (in that order), the overhead introduced by our library is small for normal enclave operations.

\begin{figure}[t]
	\begin{center}
    \includegraphics[width=1\columnwidth,trim={0 0 10mm 12mm},clip]{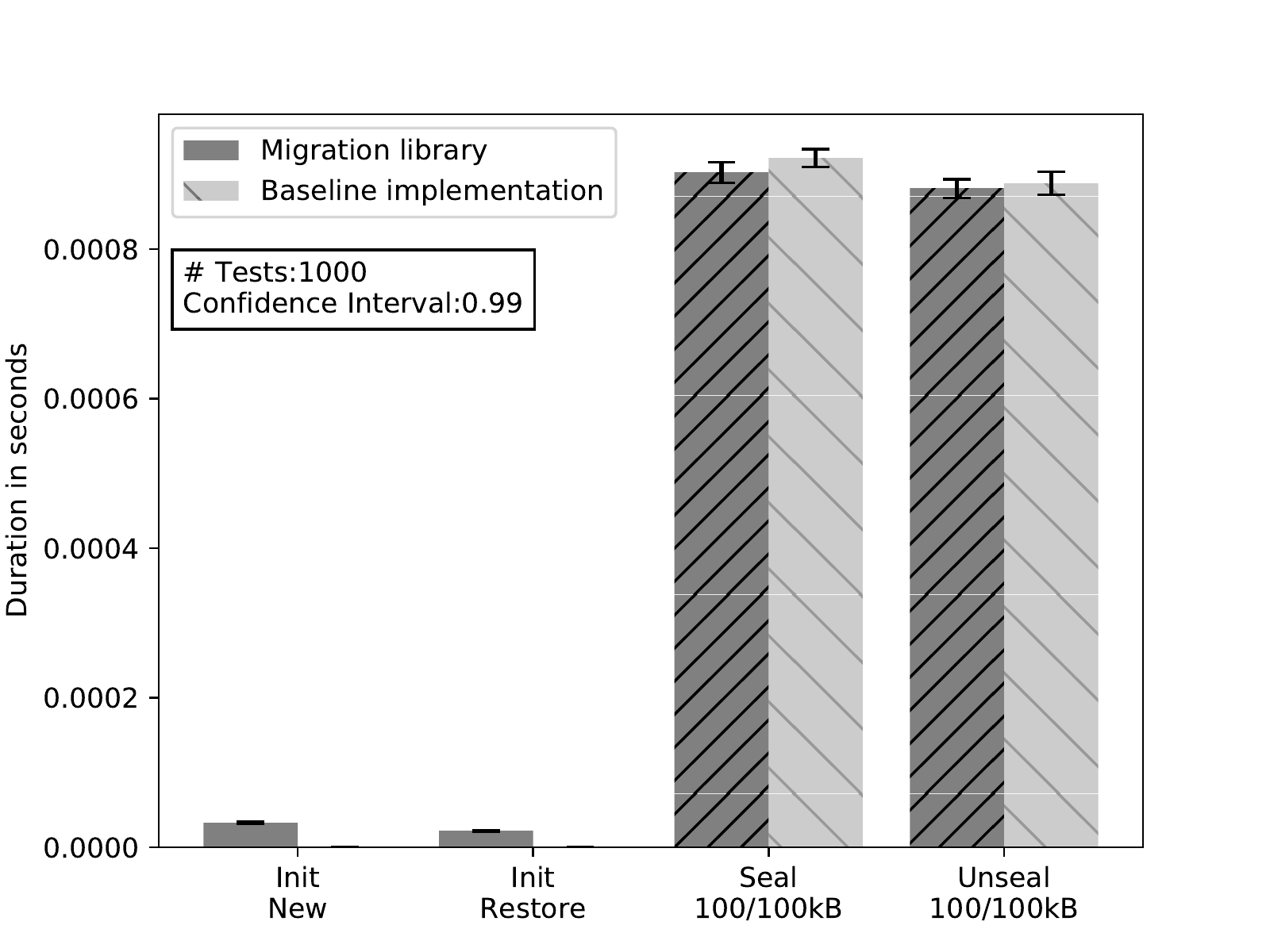}
    \caption{Average duration of initialization and sealing operations.}
    \label{fig:perf-seal}
	\end{center}
\end{figure}

Figure~\ref{fig:perf-seal} shows the measurements for the sealing and unsealing operations, as well as for the initialization of the library.
The migratable sealing operations are actually \emph{slightly faster} than their standard SGX counterparts because the MSK is already available from the library buffer.
In contrast, the standard SGX sealing operations have to perform an SGX EGETKEY operation.
The initialization of the Migration Library is very fast as it only generates a key and initializes the arrays when creating a new buffer, and only has to unseal the data when it is reloaded.
Since this is only done once during the lifetime of the enclave, the initialization time is negligible.

Additionally we measured the overhead that migrating an enclave introduces on top of VM migration.
We migrated an enclave 1000 times and calculated the average time of one migration.
The extra time for local attestation, communicating with ME and sending over the sealed data is 0.47 ($\pm$0.035) seconds. 
Since migrating the VM usually takes in order of seconds~\cite{fast-migration-of-vms}, the overhead of migrating and enclave is small by comparison.

\subsection{Usability Evaluation}

We argue qualitatively that the Migration Library is easy to use for enclave developers.
Initializing the Migration Library requires only one function call after the enclave has been started.
Similarly, initiating the migration process also only requires a single function call.
For the functions provided by the Migration Library, minimal effort is required to switch from the non-migratable versions to the migratable versions.
For sealing, only the function name changes as the other function parameters are identical to the standard SGX Library functions.
For the monotonic counter operations, the developer only has to change the function name and switch from using the SGX UUIDs to the counter\_id that is assigned by the Migration Library.
We argue that this is a reasonable low effort to make an enclave migratable and conclude that our migration system meets the defined usability goal.

\section{Future Work}
\label{sec:future}

We identify several directions for the future work.
Firstly, our proposed framework requires the VM migration to become non-transparent. 
In particular, not only the VM but also the application containing the enclave must be notified of the migration.
It is necessary to modify the enclave and explicitly register the application with the Migration Enclave.
On one hand, this allows a clear separation between migratable and non-migratable enclaves, as well as migratable and non-migratable objects within the enclave.
But on the other hand, this is in contrast to the common practice of VM migration being transparent.
Thus in some circumstances it is desirable to be able to transparently migrate unmodified enclaves.

Secondly, we focused on migrating the persistent state of the enclave.
Migrating the data memory of the enclave is an orthogonal challenge that can be solved e.g., using the mechanism proposed by Gu et al.~\cite{live-sgx-migration-dsn}.
Combining the two approaches would lead to a possibility to migrate enclaves without the need to stop and restart them.
However, we were not able to integrate the mechanism by Gu et al.~\cite{live-sgx-migration-dsn} into our system, because it was not designed to target the standard SGX SDK, and the source code is not available.

\section{Related Work}
\label{sec:related_work}

\subsection{Hardware security in the cloud}
\label{sec:related-cloud}

\changed{Although the use of hardware security technologies in the cloud is a relatively new field, there have already been several examples of their benefits in this setting.}
For example, Amazon's CloudHSM\footnote{\url{https://aws.amazon.com/cloudhsm/}} provides the functionality of a traditional hardware security module to VMs in the cloud.
In terms of trusted execution environments, Intel has published a list of recent research efforts using SGX, many of which target cloud environments.\footnote{\url{https://software.intel.com/en-us/sgx/academic-research}}
Ohrimenko et al.~\cite{oblivious-learning-usenix} provide a framework for privacy-preserving machine learning in a multi-party setting.
Using SGX, a machine learning algorithm hosted in the cloud could be provided with privacy-sensitive data from multiple parties.
Tamrakar et al.~\cite{circle-game-asiaccs} used both SGX and another type of trusted execution environment, ARM TrustZone, in a cloud setting to provide a scalable private membership test system (e.g., for cloud-based malware checking). 
In particular, they investigated how to avoid leaking data through memory access patterns.
The Signal messaging service recently announced that they are using a similar SGX-based approach on their servers to support private contact discovery.\footnote{\url{https://signal.org/blog/private-contact-discovery/}}

Zheng et al.~\cite{opaque-nsdi} described Opaque, a distributed data analytics platform, which uses SGX enclaves to support a wide range of data queries while ensuring strong security guarantees.
In the VC3 system~\cite{vc3}, SGX enclaves are used to execute a MapReduce protocol securely on untrusted infrastructure. 
In order to authenticate the machines that belong to the cloud provider, VC3 introduces the idea of a Cloud quoting enclave.
We propose a similar solution for authenticating servers owned by the cloud provider.

\changed{Matetic et al.~\cite{rote-sgx} specifically address the issue of rollback protection in TEEs.
They argue that the hardware-based monotonic counters available in SGX suffer several disadvantages, including rate-limiting and wear-out.
They propose \emph{ROTE}, a system for maintaining virtual counters using consensus among a group of SGX enclaves running on different physical machines.
A migratable enclave that uses ROTE would not need to migrate monotonic counters, but would still require a mechanism to securely migrate the keys it uses to identify itself to the ROTE system.}

\subsection{SGX migration}
\label{sec:related-migration}
The first attempt to address the problem of migrating an SGX-enabled VM was presented by Park et al.~\cite{live-sgx-migration-services}.
They identified the central challenge of copying the enclave's memory to the destination machine, which naivly appears to violate the SGX security guarantees (i.e., the source enclave's data become accessible outside the source enclave).
They proposed a conceptual solution that requires a new SGX hardware instruction.
This instruction would be use to agree on a \emph{live migration key} (LMK) between two SGX-enabled machines, and then securely transfer the contents of an application enclave from the source to the destination, encrypted using the LMK. 
One advantage of this hardware-based approach is that the migration occurs transparently to the migrating enclave, and no software changes to the enclave are required.
However, the significant disadvantage is that this would require extensive hardware changes.
This proposal has therefore not been evaluated.

Gu et al.~\cite{live-sgx-migration-dsn} presented the current state-of-the-art software-only framework to enable migration of VMs containing SGX enclaves.
They focus on the challenge of migrating the data memory of an enclave and their approach is to add a library to the enclave to support this.
Their library performs remote attestation and key agreement with the destination enclave.
It then re-encrypts the memory pages of the source enclave and writes out the resulting encrypted data outside the enclave's memory.
This can then be migrated with the VM and input to the equivalent library in the destination enclave.
Apart from successfully migrating the enclave's memory, they also considered issues of multithreading, since the enclave may be busy executing an ECALL when the migration takes place.
Their solution is to add a control thread to the enclave that will be notified of the migration.
This control thread then pauses the execution of the enclave by spin-locking all worker threads.
Although this software-only solution cannot support transparent migration, there are various ways in which the enclave can be notified about the migration.
\changed{One constraint of migrating an enclave's data memory via a software-only solution is that all data memory on the source enclave must be made readable by the migration functionality within the enclave.
The authors have presented an extensive performance evaluation of their solution, but have not released the source code.}

Our solution builds upon some the ideas presented in previous work (e.g., the migration enclave concept from Park et al.~\cite{live-sgx-migration-services} and the migration library idea from Gu et al.~\cite{live-sgx-migration-dsn}).
However, neither of these solutions consider enclaves with persistent state, leading to the potential attacks described in Section~\ref{sec:threats}.

\raggedbottom

\section{Conclusion}
\label{sec:conclusion}

\changed{Hardware-based security technologies, like Intel SGX, can provide strong security guarantees in a cloud computing setting. 
However, in order to use SGX in a cloud environment, it must be possible to migrate enclaves between physical machines.
While previous work showed how to migrate the data memory of enclaves, it did not consider migration of enclaves with persistent state (i.e., sealed data or monotonic counters). 
In this paper, we propose, design, and implement a framework to enable migration of such enclaves while maintaining the same functionality and security guarantees of non-migratable SGX enclaves. 
Our framework, which can be integrated with previous solutions, consists of a Migration Library that is integrated into each migratable enclave, and a Migration Enclave running in the management VM on each physical machine.
Our proof of concept implementation overcomes several practical challenges, such as accessing architectural enclaves from VMs and limiting the migration of enclaves to authorized machines, whilst incurring low performance overhead (less than 12.3\%) and minimal additional effort for enclave developers.}

\changed{As future work, we plan to investigate how enclave providers can provision customized \emph{migration policies} with their enclaves.
For example, a migration policy could specify minimum computational requirements of a destination machine, or ensure that a particular enclave is not migrated outside a specified geographic region.
These policies would be enforced by the Migration Enclave and the mutual local attestation between the enclave and the Migration Enclave provides assurance that they will be enforced correctly.}

\changed{As discussed in Section~\ref{sec:introduction}, fully transparent migration of enclave is not possible with current Intel SGX hardware.
In our framework, each application that uses enclaves must be notified of the migration and must call the \texttt{migrate()} function provided by our Migration Library.
However, it may be possible to perform the migration \emph{semi-transparently} by having the hypervisor or management VM locate and call the \texttt{migrate()} function of all enclaves associated with a particular VM.
The migration process will then take place as described in this paper, but will essentially be transparent to the applications and OS of the guest VM.}

\ifdefined\shownames
\fi

\raggedbottom
\raggedright

\bibliographystyle{IEEEtran}
\bibliography{sgx-migration}

\balance

\end{document}